\documentclass[10pt,conference]{IEEEtran}

\usepackage[top=0.8in, bottom=1in, left=0.635in, right=0.635in]{geometry}
\usepackage{cite}
\usepackage{amsmath,amssymb,amsfonts}
\usepackage[flushleft]{threeparttable}
\usepackage{array,booktabs,makecell}
\usepackage{algorithmic}
\usepackage{graphicx}
\usepackage{textcomp}
\usepackage{float}
\usepackage{mathtools}
\usepackage{xcolor}
\usepackage{diagbox}
\usepackage[bookmarks=false]{hyperref}
\usepackage[caption=false,font=footnotesize]{subfig}
\usepackage{siunitx}
\usepackage{multirow}
\usepackage{multicol}
\usepackage{tabularx}
\def\BibTeX{{\rm B\kern-.05em{\sc i\kern-.025em b}\kern-.08em
    T\kern-.1667em\lower.7ex\hbox{E}\kern-.125emX}}

\begin{document}

\title{
A Deep Neural Network Codebook Approach for Near-Field Nulling Control Beam Focusing
}

\author{\IEEEauthorblockN{Mohammadhossein Karimi, Yuanzhe Gong, Tho Le-Ngoc}
\IEEEauthorblockA{{Department of Electrical and Computer Engineering, McGill University, Montreal, QC, Canada} \\
Email: mohammad.karimirozveh@mail.mcgill.ca, yuanzhe.gong@mail.mcgill.ca, tho.le-ngoc@mcgill.ca}
}

 \maketitle
 
\begin{abstract}
This paper proposes a deep neural network (DNN) codebook approach for multi-user interference (MUI) mitigation in extremely large multiple-input multiple-output (XL-MIMO) systems operating in the near-field region. Unlike existing DNN-based nulling control beamforming (NCBF) methods that face scalability and complexity challenges, the proposed framework partitions the Fresnel region using correlation-based sampling and assigns a lightweight fully connected DNN model to each subsection. 
Each model is trained on beamforming weights generated using the linearly constrained minimum variance (LCMV) method, enabling accurate prediction of nulling control beam-focusing weights that simultaneously optimize the desired signal strength and suppress potential interference for both collinear and non-collinear user configurations.
Simulation results show that the trained models achieve average phase and magnitude prediction errors of 0.085 radians and 0.52~dB, respectively, across 75 sample subsections. 
Full-wave simulations in Ansys HFSS further demonstrate that the proposed DNN codebook achieves interference suppression better than 31.64~dB, with a performance gap within 2~dB of the LCMV method, thereby validating its effectiveness in mitigating MUI while reducing computational complexity.
\end{abstract}


\section{Introduction}
\label{sec:intro}
With the rapid growth in data transmission demands and the rise of technologies such as the Internet of Things (IoT), spectral efficiency has become more critical than ever in next-generation wireless networks \cite{agiwal2016next}. In sixth-generation (6G) wireless networks, extremely large multiple-input multiple-output (XL-MIMO) systems have emerged as a key enabling technology, significantly enhancing network capacity and reliability \cite{cui2022near, cui2022channel}. By incorporating a large number of antenna elements, XL-MIMO systems can form highly directional beams, effectively minimizing interference and enabling high-rate data transmission. Their increased spatial multiplexing capability also facilitates the simultaneous service of multiple users, thereby improving spectral efficiency. 

Employing a large number of antenna elements and expanding the antenna aperture leads to a quadratic increase in the Rayleigh distance, thereby enlarging the radiative near-field region \cite{zhang2022beam}. This highlights the growing importance of studying near-field communication in XL-MIMO systems. In the near-field regime, the spherical wavefront of the transmitted electromagnetic (EM) waves, unlike the planar wavefronts assumed in far-field scenarios, enables location-based beam focusing. This allows for precise control of the beamforming gain in both angular and distance domains, thereby enhancing the spatial isolation in multi-user (MU) communication systems \cite{selvan2017fraunhofer, cui2022near}. 
By leveraging this property of near-field communication, \cite{wu2023multiple} introduces the concept of location division multiple access (LDMA), which utilizes the ability to focus signals at specific locations with minimal leakage. The study also proposes a spherical-domain codebook approach for efficient beam training to support the implementation of LDMA. Furthermore, \cite{zhang2022beam} presents a near-field channel model and proposes a sum-rate optimization framework for beam focusing, applicable to digital, hybrid, and metasurface antenna architectures.
Moreover, the work presented in \cite{gong2025near} investigates various near-field nulling control beam focusing methods, including linearly constrained minimum variance (LCMV), heuristic optimization, and deep neural networks (DNN), for designing RF-stage beamformers aimed at multi-user interference (MUI) suppression.

However, as the array size increases, beam-focusing optimization algorithms become increasingly computationally intensive due to the larger matrix dimensions and the rapid growth in the number of possible user location combinations.
To address this, various studies \cite{liu2024near,zhang2023deep,gong2025near,liu2022deep,monemi20246g,mallioras2022novel,jiang2023near,karimi2025dnn} have proposed machine learning-based solutions for beamforming and beam training. In \cite{liu2024near}, a GNN-based beam training scheme for XL-MIMO systems adopts near-field channel models and leverages channel correlation among adjacent users to form a near-field codebook, aiming to reduce pilot overhead and improve beam training accuracy by mapping far-field codebooks to near-field users. Similarly, \cite{liu2022deep} introduces a CNN-based scheme that generates near-field codewords from far-field codebooks and incorporates a beam testing phase for higher accuracy. 
Lastly, \cite{monemi20246g} proposes smart spot beam focusing (SBF) using ELPM and deep reinforcement learning with the TD3 algorithm to achieve scalable, CSI-independent 3D near-field beam focusing.
Most existing NN-based approaches \cite{liu2024near,zhang2023deep,liu2022deep,monemi20246g,mallioras2022novel,jiang2023near} emphasize boosting the desired signal strength through constructive near-field beam focusing at the intended user locations; however, they overlook MUI suppression, which has a considerable impact on multi-user performance.
In \cite{mallioras2022novel}, the authors present a fully connected DNN trained on a dataset of null-steering and beamforming weights, generated using the minimum variance distortionless response (MVDR) method, to mitigate MUI. However, the study focuses exclusively on far-field scenarios with a planar wavefront model and a limited array aperture size. Effective MUI suppression through spatial isolation with spherical wavefront characteristics, achieved via nulling control beam focusing, remains insufficient, particularly in near-field MU scenarios where XL antenna arrays can accommodate a larger number of users and enable collinear MUI mitigation.
Furthermore, the performance evaluation of these existing NN-based codebooks is largely based on the assumption of ideal theoretical array radiation characteristics. Full-wave simulations that account for realistic heterogeneous element radiation patterns and inter-element mutual coupling are still required to provide more insightful illustrations of XL-array near-field beam patterns under the proposed beam-focusing algorithms.

Although recent studies \cite{gong2025near,karimi2025dnn} have investigated DNN-based solutions for near-field nulling control beamforming (NCBF), the rapidly increasing number of intended and interference user location combinations forces existing models to either limit their scope to collinear MU scenarios \cite{gong2025near} or scale to excessively large architectures \cite{karimi2025dnn} in order to maintain prediction accuracy.
None of the existing studies addresses the scalability and complexity challenges of DNN models for MUI suppression in near-field communication. 
Moreover, the non-uniform correlation variation across different distances from the array in the radiative near-field region presents an opportunity to construct a training set that captures user channel characteristics more efficiently.

In light of insights from the aforementioned studies, we propose a DNN codebook approach consisting of a library of fully connected DNN models, each dedicated to a specific subsection of the coverage area within the Fresnel region. Each model is trained on an extensive dataset tailored to its respective scenario, with ground truth values generated using the near-field LCMV implementation presented in \cite{gong2025near}. This approach reduces the complexity of individual DNN models, improving scalability and reducing inference time, as existing DNN-based solutions are often hindered by these challenges. Full-wave simulations conducted in Ansys HFSS, which account for realistic element radiation patterns and EM interactions, confirm the effectiveness of the DNN codebook method in suppressing interference in multi-user scenarios.
The remainder of this paper is organized as follows: Section II introduces the near-field phased array system model and details the NCBF algorithms. Section III describes the DNN codebook structure, including the channel correlation-based sampling strategy, DNN model architecture, training and evaluation criteria, and dataset generation process. Section IV presents results on the training and testing losses of the DNN models, along with full-wave simulation results carried out in Ansys HFSS. Finally, Section V concludes the paper and outlines future research.

\section{System Model}
\label{sec:system-model}
\subsection{Near-Field Region}
\label{near-field-region}
The near‐field region of an antenna is typically defined as the space bounded by the Fresnel and Rayleigh distances, which themselves are determined by the maximum phase misalignment between the true EM wavefront and the simplified far‐field or near‐field approximations. the Rayleigh distance ($d_{\mathrm{R}}$) and Fresnel distance ($d_{\mathrm{F}}$) \cite{zhang2022beam} are given by
\begin{equation}
    \label{eq:ray_dist}
    d_{\mathrm{R}} = \frac{2D^2}{\lambda}, \hspace{5mm}
    d_{\mathrm{F}} = \sqrt[3]{\frac{D^4}{8\lambda}},   
\end{equation}
where \(D\) is the largest antenna aperture size and \(\lambda\) is the wavelength. 
Based on wave propagation characteristics, the near-field can be further divided into two sub-regions: (1) the reactive near-field, for distances shorter than $d_{\mathrm{F}}$ and (2) the radiative near-field, for distances between $d_{\mathrm{F}}$ and $d_{\mathrm{R}}$. In the reactive near-field region, evanescent EM waves dominate, and most of the EM energy resonates in a store-and-release mode close to the antenna array, preventing feasible propagation. Within the radiative near-field region, the energy of EM waves can be focused, enabling propagation and near-field beam focusing. In contrast to the far-field region, where EM waves are assumed to have planar wavefronts and phase differences of arriving signals are attributed solely to the direction of arrival (DoA), in the near-field region, arriving EM waves exhibit spherical wavefronts, and phase variations depend on both the relative distance to the observation point and the DoA.

\subsection{Near-field Channel Model}
\label{sec:channel-model}
    Consider an observation point at position $\boldsymbol{p}$, described by spherical coordinates $(\psi,\theta,r)$. This point has a line-of-sight (LoS) link to an $N$-element antenna array and is located in the array’s near field. The antenna array response vector can then be expressed as:

    \begin{equation}
        \label{eq:channel-model}
        \boldsymbol{h}(\boldsymbol{p}) = \left[g_1 e^{j\beta\rho_1},\, g_2 e^{j\beta\rho_2},\, \dots,\, g_N e^{j\beta\rho_N}\right]^T,
    \end{equation}
    where $g_{n}$ and $\rho_{n}$ represent the amplitude scaling factor and the path length difference between the observation point and the $n$-th antenna element, respectively, and $\beta = 2\pi/\lambda$ is the wave propagation constant for wavelength $\lambda$. Under the assumption of isotropic antenna radiation patterns and LoS propagation, where $g_{n} = \lambda/(4\pi r_{n})$ and $\rho_{n} = r_{n}$, the near-field array response vector becomes
    \begin{equation}\label{eq:channel-model-extended}
    \boldsymbol{h}(\boldsymbol{p}) = \left[\frac{\lambda}{4\pi r_1}e^{j\beta r_1}, \frac{\lambda}{4\pi r_2}e^{j\beta r_2}, \dots, \frac{\lambda}{4\pi r_N}e^{j\beta r_N}\right]^T,
    \end{equation}
where $r_{n}$ denotes the distance from the observation point to the $n$-th antenna element.



    
\subsection{Multi-User Interference Suppression and LCMV}
\label{mui-suppression-lcmv}
    By aligning received signal phases, the maximum-directivity beamformer maximizes the target user’s beam-focusing gain. However, its neglect of MUI suppression usually results in a poor signal-to-interference-plus-noise ratio (SINR) in interference-limited multi-user communication systems.

    Consider a base station serving $K$ uplink users simultaneously. The $i$-th subarray, composed of $N$ antenna elements, serves the user at $\boldsymbol{p}_i$, while the remaining $K-1$ users located at $\boldsymbol{p}_k$ for $k \in \{1,\dots,i-1,i+1,\dots,K\}$ act as interferers. After applying the RF beamformer, the signal received in the $i$-th subarray is

    \begin{equation}
        \label{eq:mui-signal-model}
        y = \underbrace{\boldsymbol{w}_i^T\boldsymbol{h}(\boldsymbol{p}_i)s_i}_{\text{Desired Signal}} + \underbrace{\sum_{k \neq i}\boldsymbol{w}_i^T\boldsymbol{h}(\boldsymbol{p}_k)s_{k}}_{\text {Multi-user Interference}}  + \underbrace{\boldsymbol{w}_i^T \boldsymbol{n}}_{\text {Noise}}.
    \end{equation}

    Here, $\boldsymbol{w}_i\in\mathbb{C}^{N\times1}$ is the RF beamforming vector applied to the $i$-th subarray, $s_i$ and $s_k$ are the symbols transmitted by the $i$-th (desired) and $k$-th (interfering) users, respectively, and $\boldsymbol{n}\in\mathbb{C}^{N\times1}$ denotes additive noise.

    The LCMV beamformer is a widely adopted NCBF technique \cite{van2002optimum}. By enforcing the array response to follow a predefined beam-focusing gain vector while minimizing interference power, the LCMV beamformer places deep radiation nulls at the $K-1$ interference user locations $\boldsymbol{p}_k$ ($k\neq i$) and maintains the desired gain for the desired user located at $\boldsymbol{p}_i$. The relationship between the constraint matrix and the desired response vector is

\begin{equation}
    \boldsymbol{C}^H\boldsymbol{w}_{\mathrm{LCMV}} = \boldsymbol{d},
    \label{eq:condition_matrix}
\end{equation}
where $\boldsymbol{C} = [\boldsymbol{h}(\boldsymbol{p}_1), \dots, \boldsymbol{h}(\boldsymbol{p}_K)]\in \mathbb{C}^{N\times K}$ is the array response matrix for all $K$ users, and $\boldsymbol{d}\in \mathbb{R}^{K\times1}$ is the desired gain vector with $d_i=1$ and $d_k=0$ for $k\neq i$. Given the signal covariance matrix $\mathbf{R}$, the optimal weight vector for an $N$-element array is obtained in closed form as \cite{van2002optimum,gong2024perturbation}:

\begin{equation}
    \boldsymbol{w}_{\mathrm{LCMV}}(\boldsymbol{p}_1,\dots,\boldsymbol{p}_K)
= \mathbf{R}^{-1}\,\boldsymbol{C}\,\bigl(\boldsymbol{C}^H\mathbf{R}^{-1}\boldsymbol{C}\bigr)^{-1}\,\boldsymbol{d}.
\label{eq:lcmv_formula}
\end{equation}

\section{Codebook Design}
\label{codebook-design}
Studies in \cite{gong2025near,mallioras2022novel} demonstrated that deep neural networks can effectively perform nulling‐control beamforming in both near- and far-field scenarios. However, extending the near-field region with extra-large antenna arrays and accommodating both collinear and non-collinear user positions—rather than relying solely on angular beamforming in the far field—greatly expands the spatial domain. Consequently, training a single model to predict beamforming weights over this extended coverage area becomes computationally prohibitive.

Dividing the designated coverage area into smaller sections enables the use of separate, lightweight models for each subsection, thereby simplifying the prediction of beam-focusing weights. In this section, we present the codebook design procedure, which includes selecting subsections via correlation-based sampling and describing the architecture of the DNN-based codebook.

\subsection{Channel Correlation-based Sampling}
\label{correlation-sampling}
To partition the near-field region of an $N$-element antenna into smaller subsections, we sample the channel steering vectors over both angular and radial domains, thereby maintaining a constant correlation between adjacent samples. 
Because beam resolution, as well as focusing and nulling capabilities, are primarily determined by the number of array elements in the considered plane, a uniform linear array (ULA) sub-array configuration with an effective steering angle \(\theta\) is adopted in this study.
With the correlation-based sampling method, the radial distance samples along the $\theta$ angle are given by
\begin{equation}
    r_{s}(\theta) = \frac{1}{s}\,\frac{N^{2}d^{2}}{2\lambda\,\beta_{\Delta}}\,\cos\theta,\quad s = 0,1,2,\dots,
    \label{eq:distance_samples}
\end{equation}
where $s$ is the sampling index, $d$ is the element spacing of the ULA, and $\beta_{\Delta}$ represents the value corresponding to the desired correlation level, derived from the Fresnel approximation of correlation between adjacent samples \cite{wu2023multiple}.

Furthermore, To ensure orthogonality among angular samples, the angles are chosen as
\begin{equation}
    \theta_n = \cos^{-1}\left({\frac{2n - N + 1}{N}}\right), \quad n = 0, 1, \dots, N-1,
    \label{angular_samples}
\end{equation}
where $n$ denotes the angular sample index.

\begin{figure}[t]
    \centering
    \includegraphics[width=0.9\linewidth]{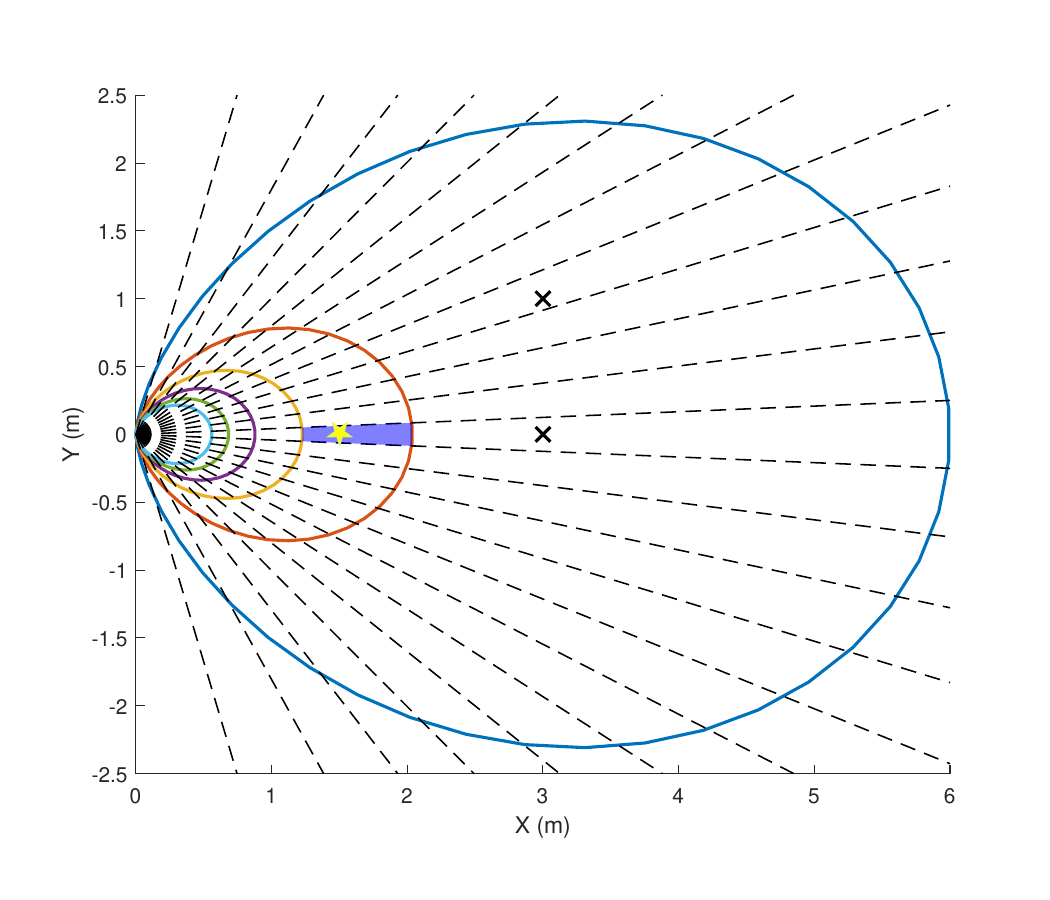}
    \vspace{-0.2cm}
    \caption{Partitioning of the near-field region, illustrating radial distance samples (solid lines) and angular samples (dashed lines), with target and interfering users placed in their respective subsections.}
    \vspace{-0.2cm}
    \label{fig:correlation-sampling}
\end{figure}

Combining angular and radial samples partitions the coverage area. Fig.~\ref{fig:correlation-sampling} illustrates the samples obtained between radial distances of 0.5 m and 6 m for a 1$\times$24 ULA placed along the y-axis and operating at 3.5 GHz, where the Rayleigh distance is approximately 20 m. As the distance from the array increases, the spacing between samples also grows, reflecting the reduced variation in differential phase shifts across the array and the slower change in correlation. This behavior highlights the necessity of correlation-based sampling instead of uniform radial sampling, which cannot capture this behavior and significantly degrades the efficiency of the samples.

\subsection{DNN Codebook Structure}
\begin{figure}[t]
    \centering
    \includegraphics[width=\linewidth]{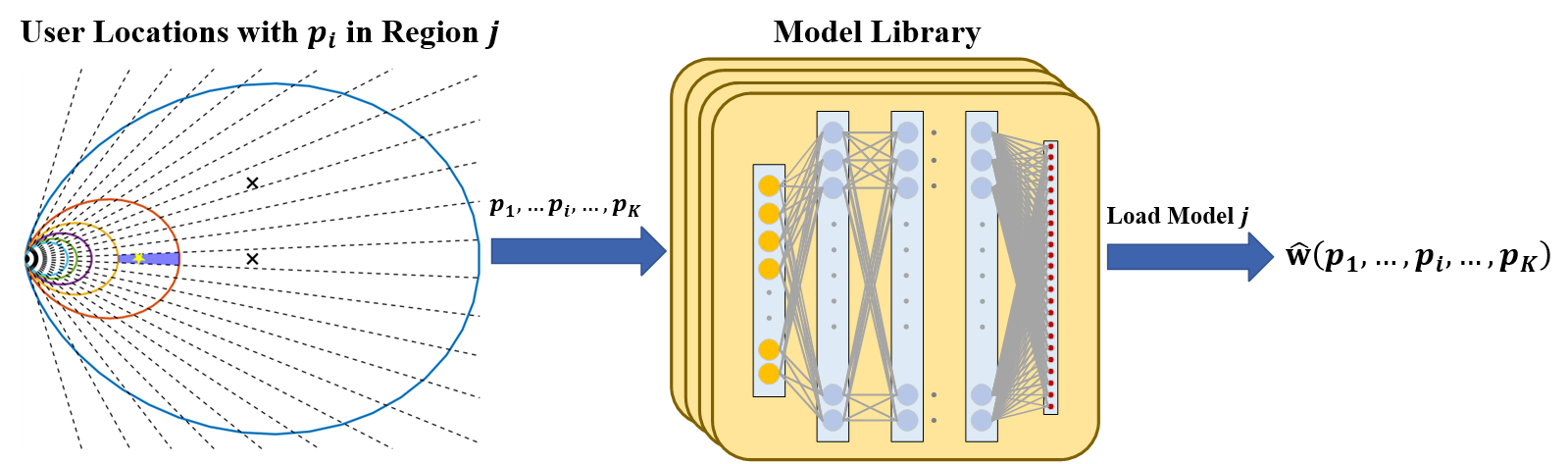}
    \caption{Structure of DNN codebook for MUI suppression.}
    \label{fig:codebook_structure}
    \vspace{-0.2cm}
\end{figure}

To balance DNN model complexity and size with their effectiveness in suppressing MUI, we employ a DNN codebook composed of multiple compact models. Each model is specialized for target users within a specific subsection of the coverage area, as defined by correlation‐based sampling. Fig.~\ref{fig:codebook_structure} illustrates the multi‐user interference suppression process using this DNN codebook.

Considering a desired user at $\boldsymbol{p}_i$ within region $j$, we first preprocess the locations of the desired and interfering users to form the input for the DNN. We then load the DNN model corresponding to region $j$ and use it to predict the beam‑focusing weights for multi‑user interference suppression. This approach requires storing one instance of the model for each subsection and training each on data reflecting scenarios with the desired user located in its respective region.

\paragraph{DNN Model Structure}
 For each region in the codebook, we employ a dual-estimator DNN architecture: two networks predict the phase and magnitude of the beam‑focusing weights separately. Each network’s input layer receives the polar coordinates of all user locations. The size of the output layer is set to the number of antenna elements, corresponding to the number of phase shifters or attenuators in the RF beamformer. The hidden‑layer dimensions are chosen to balance prediction accuracy with the model size and complexity.
\paragraph{Datasets and Preprocessing}
To train and evaluate the DNN models, we create separate training and testing datasets for each model, specifically for magnitude and phase predictions. We utilize the LCMV method to generate effective beam-focusing vectors for near-field continuous beamforming. Each dataset comprises pairs of input data and the corresponding output labels, where the inputs consist of polar coordinates indicating the locations of the desired and interfering users. These inputs are normalized during preprocessing to improve model convergence and training efficiency. The output vector for the phase-estimation model contains phase-shift values corresponding to each antenna element, with the phase of the first element shifted to zero as the reference. This adjustment structures the labels more consistently, eliminating arbitrary bias, as the beam-focusing performance of a phased array relies solely on differential phase shifts across antenna elements. For the magnitude-estimation model, the outputs consist of the magnitudes of beam-focusing weights. To ensure unit-power NCBF weights, these magnitude components derived from the LCMV algorithm are scaled by a factor of $1/\sqrt{\sum_{n=1}^{N} a_n^2}$, where $a_n$ is the magnitude of the beam-focusing weight corresponding to the $n$-th antenna element of the $N$-element array. Furthermore, the magnitude labels are converted to the dB scale to enhance numerical stability and facilitate the training process.

\paragraph{Loss Functions}
Training and evaluating the performance of the models in the prediction stage requires a well-defined measure of error between the DNN predictions and the labeled NCBF weights obtained using LCMV. For phase-estimation models, we employ the circular mean absolute error (CMAE), which computes the minimum angular distance between the predicted phase-shift vector $\hat{\boldsymbol{\phi}} = [\hat{\phi}_1, \hat{\phi}_2, \dots, \hat{\phi}_N]$ and the ground-truth phase vector from the LCMV-based NCBF weights, $\boldsymbol{\phi} = [\phi_1, \phi_2, \dots, \phi_N]$. Using CMAE, which considers the periodic nature of phases, removes ambiguities arising from linear measures of error. The CMAE between the predicted phases $\hat{\boldsymbol{\phi}}$ and the true phases $\boldsymbol{\phi}$ is calculated as:

\begin{equation}
    \label{eq:CMAE}
    J_{\phi}(\boldsymbol{\phi}, \boldsymbol{\hat{\phi}}) = \frac{1}{N} \sum_{n = 1}^{N} \min\left( \left| \hat{\phi}_n - \phi_n \right|, \, \left|2\pi - \left| \hat{\phi}_n - \phi_n \right| \right| \right).
\end{equation}

For magnitude estimation, we employ the root mean squared error (RMSE) to quantify the discrepancy between the predicted magnitude vector $\hat{\boldsymbol{a}} = [\hat{a}_1, \hat{a}_2, \dots, \hat{a}_N]$ and the true magnitude vector $\boldsymbol{a} = [a_1, a_2, \dots, a_N]$. The RMSE used to evaluate the magnitude estimation DNN model is defined as
\begin{equation}
    \label{eq:RMSE}
    J_{a}(\boldsymbol{a}, \boldsymbol{\hat{a}}) = \sqrt{\frac{1}{N} \sum_{n=1}^{N} (a_n - \hat{a}_n)^2}.
\end{equation}

\section{Illustrative Results}
In this study, we consider a $1\times24$ ULA antenna operating at 3.5\,GHz with an element spacing of 4\,cm (equivalent to $0.467\lambda$) and a Rayleigh distance of 19.75\,m. Previous works \cite{kosasih2024finite,gong2025near} have shown that near-field beam focusing for collinear users is feasible only within a limited portion of the near-field region. This limitation arises from the rapid increase in beam depth along the radial dimension, which makes it difficult to distinguish two collinear users at greater distances. To ensure the quality of our training data, we define a coverage area radially from 0.5\,m to 6\,m (approximately $0.02d_R$ to $0.3d_R$) and angularly between $-40^{\circ}$ and $40^{\circ}$. Partitioning this coverage area using correlation-based sampling with intersample correlation values of 0.7, 0.6, and 0.4 yields $M_C =$ 75, 60, and 45 subsections, respectively. Here, $M_C$ represents the number of spatial sectors in the codebook. Each subsection is associated with two DNN models (for phase and magnitude estimation), each comprising six hidden layers structured as $\left[2K, 1024, 512, 512, 256, 128, 64, N\right]$.

\subsection{Training and Evaluation of Models in the Codebook}
The DNN models for each subsection are trained on 100{,}000 samples, split 80\% for training and 20\% for testing. Training is performed for 200 epochs with a batch size of 1{,}000 to ensure stable convergence. An exponential learning rate decay with a factor of 0.97 is applied to smooth convergence and prevent divergence from the optimal solution.


\begin{figure}[t]
	\centering
	\subfloat[]{\includegraphics[width=0.49\columnwidth]{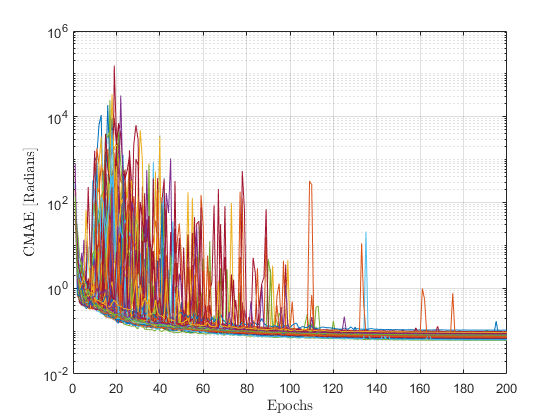}\label{fig:3-a}}\hfill
	\subfloat[]{\includegraphics[width=0.49\columnwidth]{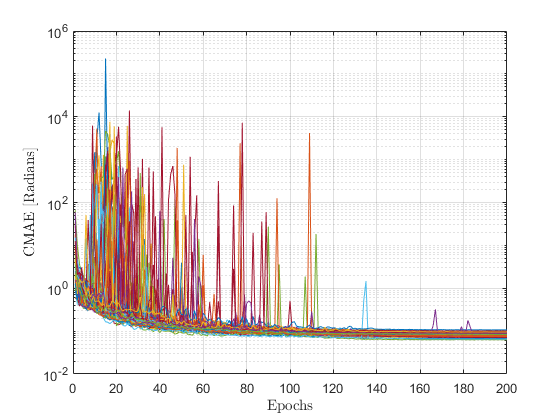}\label{fig:3-b}}\\[-10pt]
	\subfloat[]{\includegraphics[width=0.49\columnwidth]{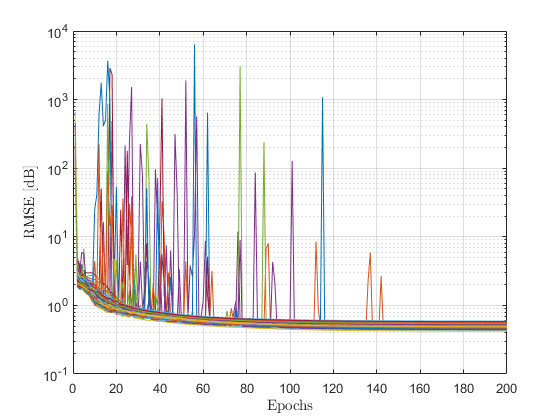}\label{fig:3-c}}\hfill
	\subfloat[]{\includegraphics[width=0.49\columnwidth]{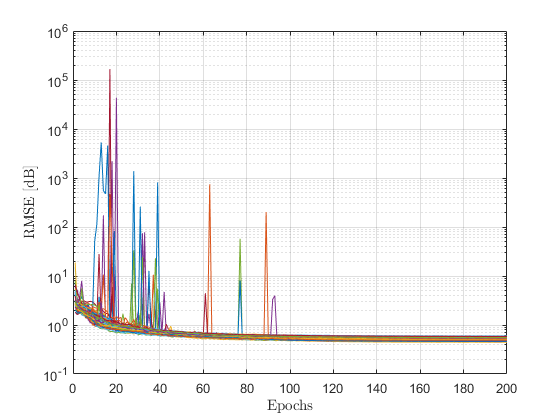}\label{fig:3-d}}
	\caption{Training and testing losses over the training process for $M_C=75$: (a–b) phase, (c–d) magnitude.}
	\label{fig:loss_plots}
\end{figure}

Fig.~\ref{fig:loss_plots} illustrates the training and testing loss curves over the training epochs for the case of $M_C = 75$. Initially, the loss curves exhibit high error levels and large fluctuations, attributed to the non-convex nature of the loss function. However, after approximately 100 epochs, the number of spikes significantly decreases, and the losses begin to converge toward a steady state. Ultimately, the training and testing losses for the phase estimation models stabilize around 0.085 radians, while the magnitude estimation models converge to approximately 0.52\,dB. The average and standard deviation (SD) of the training and testing losses for both models are summarized in TABLE~\ref{tab:loss_stats}.


\begin{table}[t]
    \centering
    \caption{Train and test loss statistics}
    \renewcommand{\arraystretch}{1}
    \scriptsize
    \begin{tabularx}{\columnwidth}{c|c|>{\centering\arraybackslash}X
                                    >{\centering\arraybackslash}X|
                                    >{\centering\arraybackslash}X
                                    >{\centering\arraybackslash}X}
        \midrule
        \multirow{2}{*}{\textbf{Codebook Size}} & \multirow{2}{*}{\textbf{Estimator}} 
        & \multicolumn{2}{c|}{\textbf{Train Loss}} 
        & \multicolumn{2}{c}{\textbf{Test Loss}} \\
        \cline{3-6}
        & & \textbf{Mean} & \textbf{SD} & \textbf{Mean} & \textbf{SD} \\
        \hline
        \multirow{2}{*}{\(\mathbf{M}_C = 75\)} 
            & Phase [rad]    & 0.083 & 8.6E-3 & 0.085 & 8.4E-3 \\
            & Magnitude [dB] & 0.504 & 0.037  & 0.518 & 0.036 \\
        \hline
        \multirow{2}{*}{\(\mathbf{M}_C = 60\)} 
            & Phase [rad]    & 0.088 & 0.02 & 0.090 & 0.02 \\
            & Magnitude [dB] & 0.511 & 0.035  & 0.525 & 0.035 \\
        \hline
        \multirow{2}{*}{\(\mathbf{M}_C = 45\)} 
            & Phase [rad]    & 0.091 & 0.015 & 0.093 & 0.015 \\
            & Magnitude [dB] & 0.521 & 0.034  & 0.537 & 0.033 \\
        \midrule
    \end{tabularx}
    \vspace{-0.5cm}
    \label{tab:loss_stats}
\end{table}

\begin{figure}[t]
	\centering
	\subfloat[]{\includegraphics[width=0.49\columnwidth]{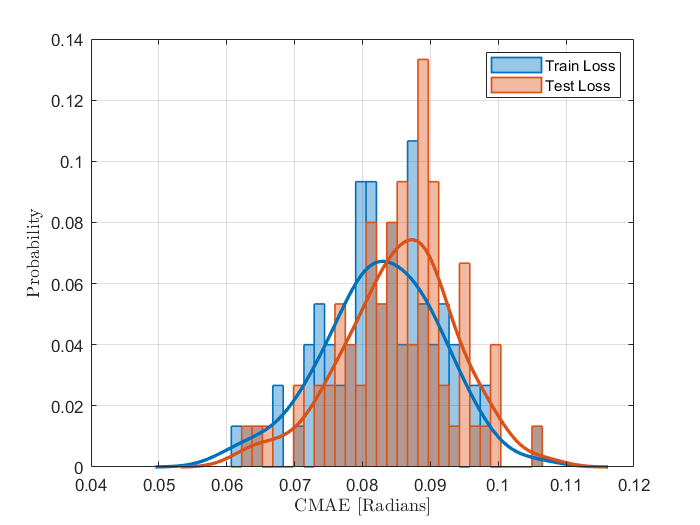}\label{fig:4-a}}\hfill
	\subfloat[]{\includegraphics[width=0.49\columnwidth]{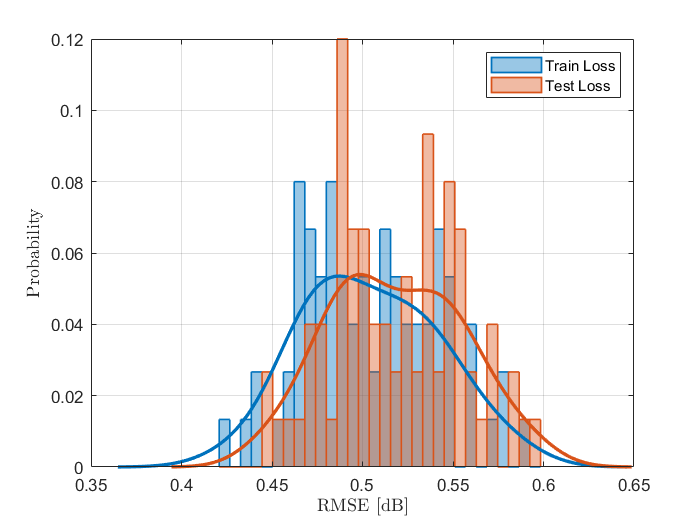}\label{fig:4-b}}
	\caption{Distribution of training and testing losses across the codebook subsections for: (a) phase estimator DNN, and (b) magnitude estimator DNN.}
	\label{fig:loss_dist}
\end{figure}

The results presented in TABLE~\ref{tab:loss_stats} present the training and test losses across codebook sectors for different codebook sizes. These results demonstrate that the trained DNN models achieve high accuracy with relatively low error on both the training and unseen test data. Furthermore, the small standard deviation values across subsections indicate consistent and uniform model performance throughout the codebook. However, when the number of sectors is reduced, both the mean and standard deviation of the training and test losses increase. The error distributions for phase and magnitude estimation with the $M_C=75$ codebook are shown in Fig.~\ref{fig:loss_dist}.


The distribution of training and testing loss values for phase estimation, shown in Fig.~\ref{fig:4-a}, indicates that the errors range approximately from 0.06 to 0.1 radians, with most subsections concentrated around the average value of 0.085 radians. Similarly, Fig.~\ref{fig:4-b} shows the distribution of losses for magnitude estimation, where the RMSE values range from 0.42 to 0.6\,dB, the majority centered on 0.5\,dB. The slight shift observed in the distribution of test errors is due to a marginally higher average test loss.

\subsection{Full-Wave Simulations}
%
%

\begin{figure*}[t]
	\centering
	\subfloat[]{\includegraphics[width=0.3\textwidth]{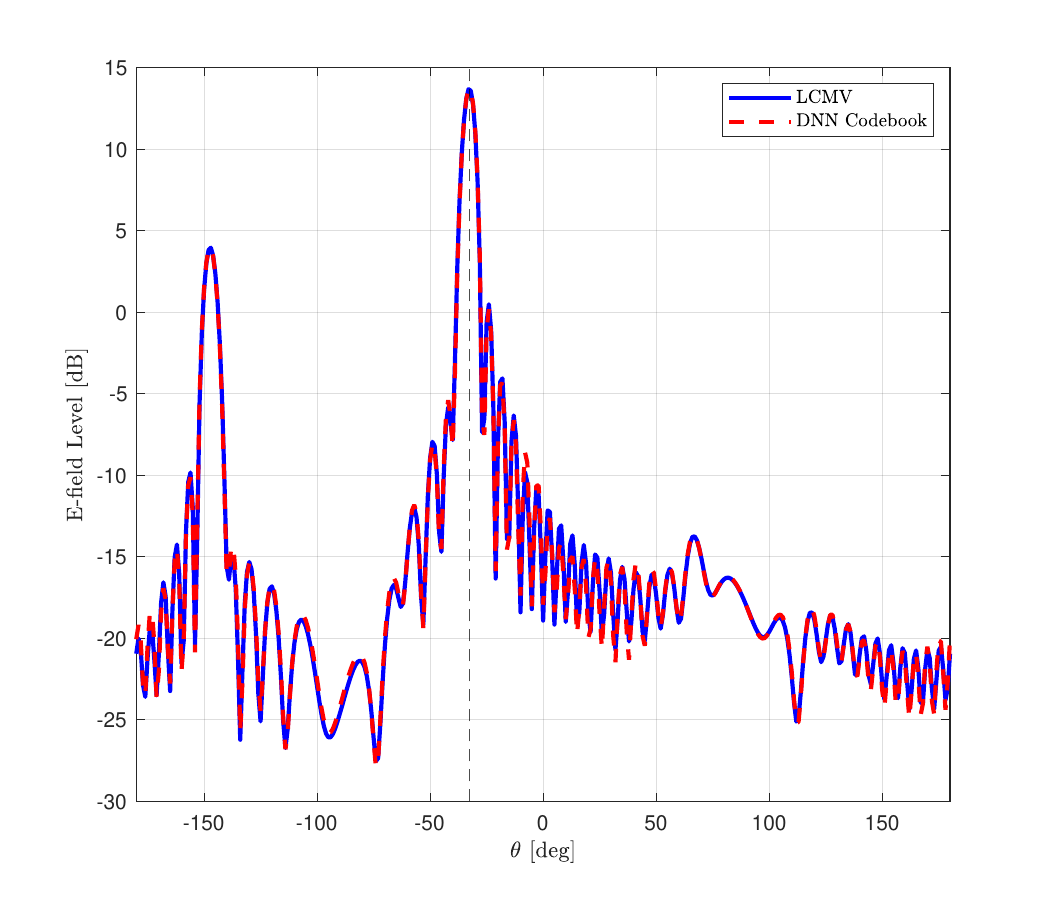}\label{fig:5-a}}\hfill
	\subfloat[]{\includegraphics[width=0.3\textwidth]{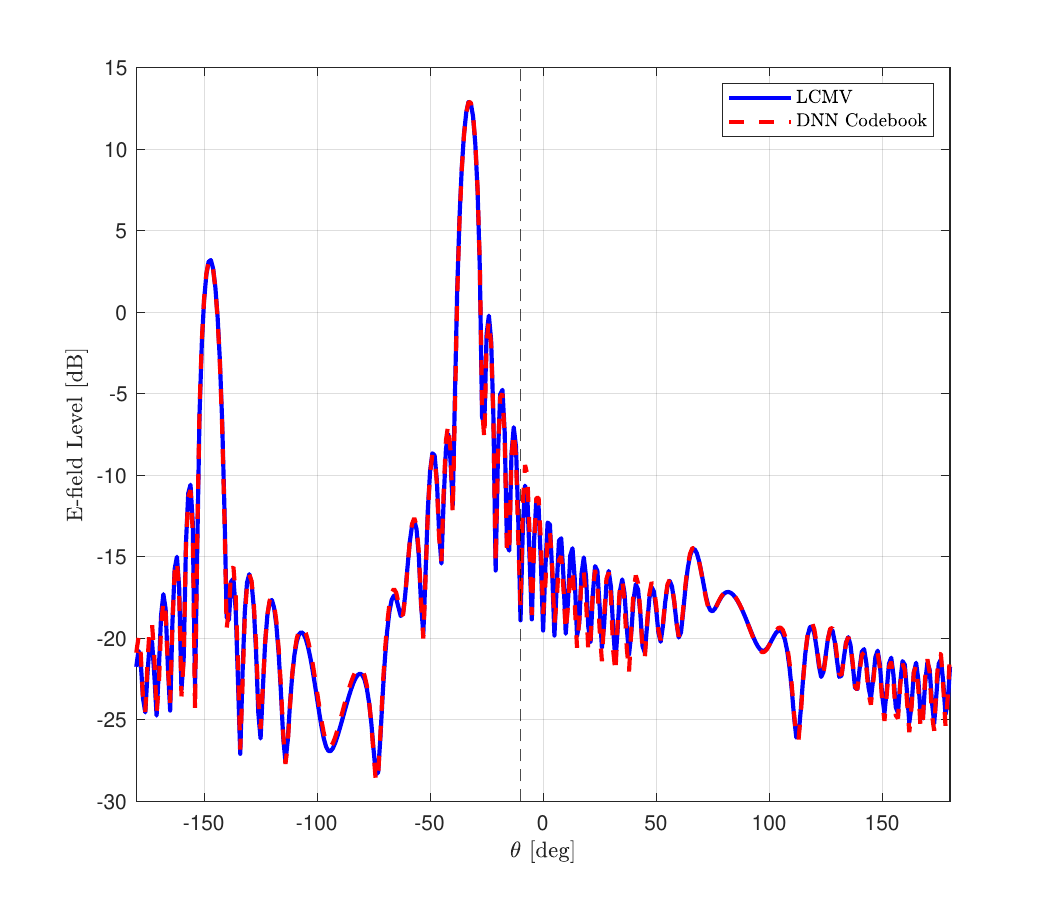}\label{fig:5-b}}\hfill
	\subfloat[]{\includegraphics[width=0.3\textwidth]{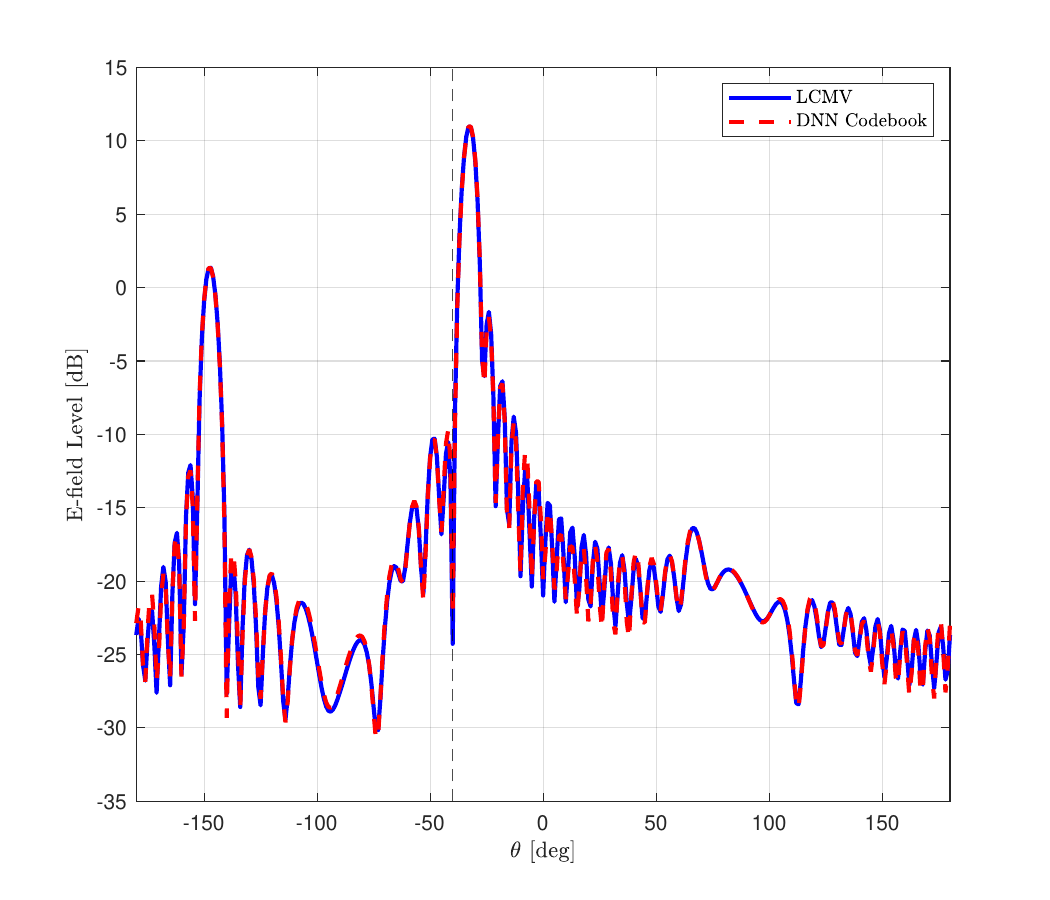}\label{fig:5-c}}\\[-10pt]
	\subfloat[]{\includegraphics[width=0.3\textwidth]{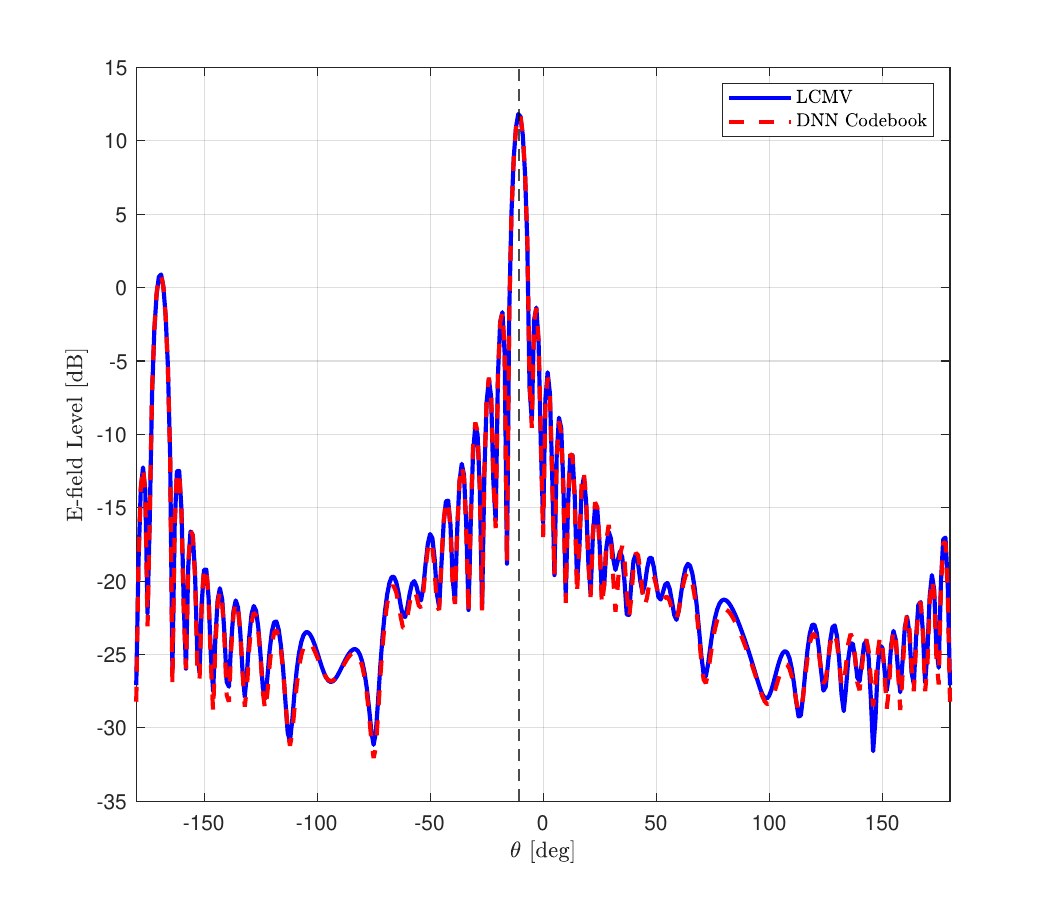}\label{fig:5-d}}\hfill
	\subfloat[]{\includegraphics[width=0.3\textwidth]{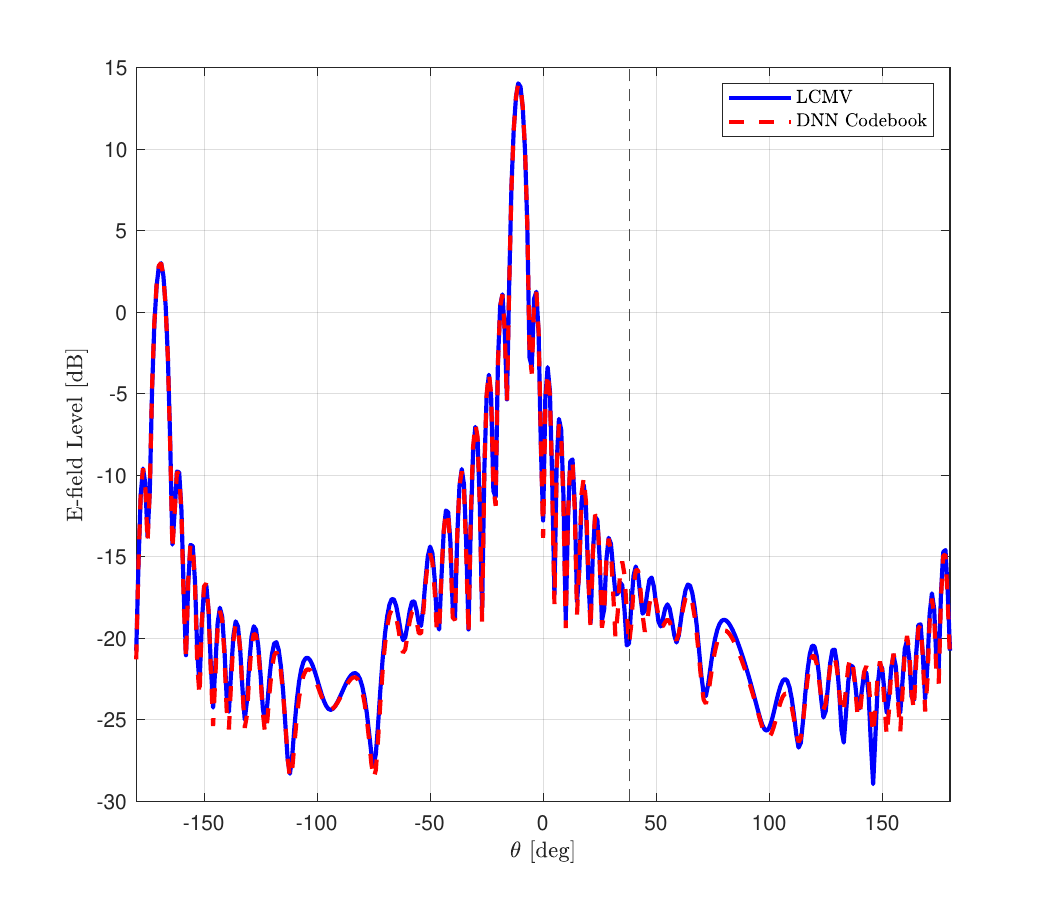}\label{fig:5-e}}\hfill
	\subfloat[]{\includegraphics[width=0.3\textwidth]{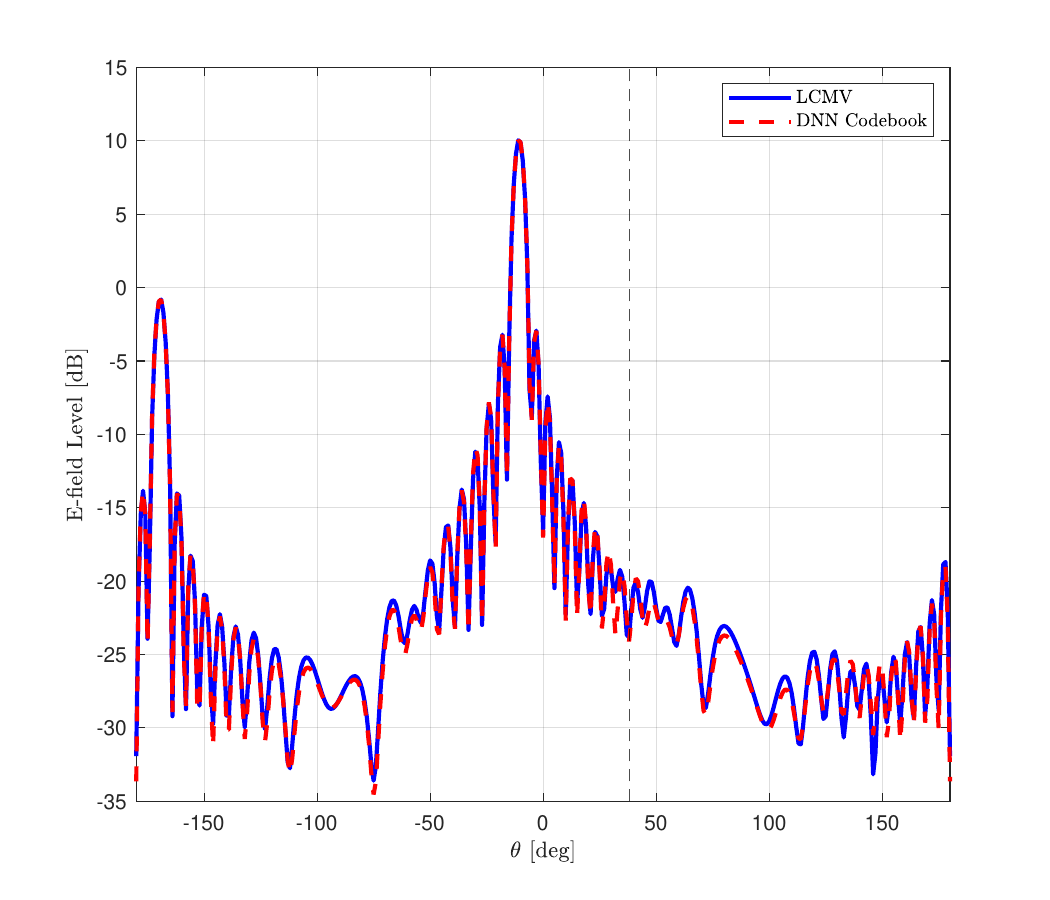}\label{fig:5-f}}

    \caption{2D cuts of Ansys HFSS simulated radiation patterns for two three-user scenarios. Scenario 1: desired user is located at (a)  $\boldsymbol{p}_1= (-32^\circ, 3.4\,\mathrm{m})$, and interfering users at (b) $\boldsymbol{p}_2= (-10^\circ, 3.7\,\mathrm{m})$ and (c) $\boldsymbol{p}_3= (-40^\circ, 4.6\,\mathrm{m})$. Scenario 2: desired user is located at (d) $\boldsymbol{p}_1= (-10^\circ, 4.75\,\mathrm{m})$, and interfering users at (e) $\boldsymbol{p}_2= (38^\circ, 3.65\,\mathrm{m})$, and (f) $\boldsymbol{p}_3= (38^\circ, 5.8\,\mathrm{m})$}
    \vspace{-0.3cm}
    \label{fig:beampatterns}
\end{figure*}

To further evaluate the performance of the proposed DNN codebook in MUI suppression and nulling controls, a $1\times24$ metamaterial-based patch antenna is employed and simulated in Ansys HFSS. Detailed information on the patch element design is presented in \cite{gong2022miniaturized}. In this setup, NCBF weights obtained from the DNN codebook are employed and compared against NCBF weights directly computed via LCMV to assess their effectiveness in MUI suppression. The full-wave simulation results of two sample three-user near-field scenarios are illustrated in Fig.~\ref{fig:beampatterns}. 

Figs.~\ref{fig:5-a}–\ref{fig:5-c} depict the first scenario, where all three users have different angular and radial coordinates. In this case, the beam patterns generated by the DNN codebook closely match those obtained using LCMV. Specifically, the DNN codebook approach achieves interference suppression levels of 31.64\,dB and 35.96\,dB between the desired and interfering users, with a maximum deviation of less than 2\,dB in the nulling level compared to LCMV.

Figs.~\ref{fig:5-d}–\ref{fig:5-f} present the second scenario, in which two users are placed collinearly. Similar to the first case, the beam patterns produced by the DNN codebook align closely with those from LCMV. A gain difference of 32.18\,dB and 35.93\,dB between the desired and interfering users is achieved, exactly matching the LCMV results.
These two scenarios demonstrate the effectiveness of the proposed DNN codebook approach in handling both collinear and non-collinear user placements in near-field communication.

\section{Conclusions}
This paper presents a deep neural network (DNN) codebook approach for multi-user interference mitigation in XL-MIMO systems. The proposed method addresses the scalability challenges of existing DNN-based solutions for near-field NCBF, which often require extensive computational resources and struggle to operate in near real-time. To reduce complexity, we apply correlation-based sampling to partition the near-field region and construct a codebook of lightweight DNN models for efficient beamforming weight prediction. The performance of the proposed approach is evaluated through both prediction error analysis and full-wave simulations in Ansys HFSS, demonstrating its effectiveness in mitigating multi-user interference in near-field communication.

\bibliographystyle{IEEEtran}
\bibliography{bibliography.bib}

\end{document}